# Investigation of helicity-dependent photocurrent at room temperature from a Fe/x-AlO$_x$/p-GaAs Schottky junction with oblique surface illumination

Ronel Christian Roca*, Nozomi Nishizawa, Kazuhiro Nishibayashi, and Hiro Munekata

*Institute of Innovative Research, Tokyo Institute of Technology, Yokohama 226-8503, Japan*

*E-mail: ronel.roca@isl.titech.ac.jp

In view of a study on spin-polarized photodiodes, the helicity-dependent photocurrent ($\Delta I$) in a Fe/γ-AlO$_x$/p-GaAs Schottky diode is measured at room temperature by illuminating a circularly polarized light beam ($\lambda$ = 785 nm) either horizontally on the cleaved sidewall or at an oblique angle on the top metal surface. The plane of incidence is fixed to be parallel to the magnetization vector of the in-plane magnetized Fe electrode. The conversion efficiency $F$, which is a relative value of $\Delta I$ with respect to the total photocurrent $I_{ph}$, is determined to be 1.0 × 10$^{-3}$ and 1.2 × 10$^{-2}$ for sidewall illumination and oblique-angle illumination, respectively. Experimental data are compared with the results of a model calculation consisting of drift-diffusion and Julliere spin-dependent tunneling transports, from which two conclusions are obtained: the model accounts fairly well for the experimental data without introducing the annihilation of spin-polarized carriers at the γ-AlO$_x$/p-GaAs interface for the oblique-angle illumination, but the model does not fully explain the relatively low $F$ in terms of the surface recombination at the cleaved sidewall in the case of sidewall illumination. Microscopic damage to the tunneling barrier at the cleaved edge would be one possible cause of the reduced $F$.

## 1. Introduction

Spin-based optoelectronic devices are unique in the sense that they utilize of the mutual conversion of spins between light and carriers, and they may add new functionalities to conventional optoelectronic devices. Some of the interesting potential applications include chiral molecule resolution[1,2] and advanced optical communication based on circularly polarized (CP) light.[3,4] In particular, the inherent association of the electron spin orientation in zinc blende and diamond crystal semiconductors, with the circular polarization of light[5,6] allows for the fabrication of devices that can natively generate and detect CP light; examples are spin-polarized light-emitting diodes (spin-LEDs)[7,8] and spin-polarized photodiodes (spin-PDs)[9].

Most previous works dealt with vertical-type spin-PDs.[3,9-16] In early works, the conversion efficiency or figure of merit $F = \Delta I/I_{ph}$ was introduced, in which $|\Delta I| = |I_{ph}(\sigma^+) - I_{ph}(\sigma^-)|$ and $I_{ph} = [|I_{ph}(\sigma^+) + I_{ph}(\sigma^-)|/2]$.[9] Here, $\sigma^+$ and $\sigma^-$ represent right and left CP light beams, respectively. On the basis of this definition, $F \approx 5\%$ at room temperature has been achieved.[9,10] In a vertical configuration, a light beam is irradiated onto the top surface of a spin-PD usually through the magnetic metal contact. It restricts the choice of magnetic materials since perpendicular magnetic anisotropy (PMA) is necessary. Besides, in the case of direct device-to-device communication with light,[3] precise chip-to-chip alignment may be required. In order to compensate for the limitations of vertical spin-PDs, lateral spin-PDs are desired, where light is irradiated onto the side or edge of a spin-PD. These devices do not require PMA and allow for simpler intrachip communications. However, works on lateral-type spin-PD have been scarce, and $F$ has remained somewhat poor ($F \approx 0.1\%$).[17]

In this work, we study the origin of the low $F$ in a lateral spin-PD using oblique-angle surface illumination. In this configuration, surface recombination on the cleaved sidewall can be avoided, which allows us to compare and experimentally assess the contribution of surface recombination. The experiment results have been presented in the International Conference on Solid State Devices and Materials (SSDM) 2016.[17] We have used p-GaAs instead of the *n-i-p* structure[18] on the basis of the inference that the depletion region in a Fe/γ-AlO$_x$/p-GaAs Schottky diode allows photogenerated electrons to drift efficiently towards the Fe/γ-AlO$_x$ electrode and experience a large change in quasi-Fermi level as compared with n-type GaAs upon illumination. Dyakonov-Perel (DP) and Bir-Aronov-Pikus (BAP) mechanisms are the dominant sources of spin relaxation at room temperature in p-GaAs.[19] We have carried out, in addition to experiments, a model calculation using drift-diffusion transport and Julliere tunneling spin transport equations in order to quantitatively assess the influence of surface recombination on the cleaved surface. We show that the model explains fairly well the experimental data without incorporating any degradation of spin-polarized carriers at the γ-AlO$_x$/p-GaAs interface for the oblique-angle illumination. For sidewall illumination, the total photocurrent is explained by the model with the surface recombination, whereas the low helicity-dependent photocurrent is not fully explained by the surface recombination. Microscopic damage to the tunneling barrier at the cleaved edge is suggested as a possible cause of the reduced helicity-dependent photocurrent.

## 2. Experimental methods

A schematic of the sample layer structure is shown in Fig. 1(a). On a p-GaAs:Zn ($N_A$ ~10$^{-19}$ cm$^3$) substrate, a 100-nm-thick epitaxial p-GaAs:Be ($N_A$ ~10$^{18}$ cm$^{-3}$) layer followed by a 1-nm crystalline γ-like AlO$_x$ (γ-AlO$_x$) layer was grown by MBE. Details of the growth process for γ-AlO$_x$ can be found elsewhere.[20] This layer acts as the tunnel barrier, which is required to circumvent the conduction mismatch problem in semiconductor spintronic devices,[12] and impedes the chemical reaction between Fe and GaAs. The top metal contact comprising a 50-nm Fe layer, followed by 5-nm Ti and 10-nm Au layers, was deposited by e-beam and Joule-heat evaporations. Furthermore, a 40-nm indium layer was deposited on the back side of the substrate as the ohmic contact. The sample was then annealed at 230 °C for 1 h in N$_2$ environment. The ferromagnetic Fe layer exhibits in-plane magnetic anisotropy with a coercive force of approximately 110 Oe [Fig. 5(b)]. Finally, the sample was cleaved into a rectangular chip with dimensions of ~1 × ~2 mm$^2$. The Schottky junction formed on the semiconductor side of γ-AlO$_x$/p-GaAs is supposed to transport the photogenerated electrons efficiently toward the Fe/γ-AlO$_x$/p-GaAs tunnel junction [Fig. 1(d)]. The width of the depletion region is estimated to be ~30 nm. We use $P_{Fe} = 0.42$ the spin polarization of Fe.[21-23]

The experimental setup is shown in Fig. 1(b). A 20 mW laser beam ($\lambda = 785$ nm) was focused on the sample surface with a spot size (FWHM) of approximately 400 μm. To generate circular polarization, a linear polarizer (LP) and a quarter-wave plate (QWP) were used. The helicity of the CP light was switched between σ$^+$ and σ$^-$ by manually rotating QWP. All measurements were carried out at room temperature using the lock-in technique with a mechanical chopper. The helicity-dependent photocurrent Δ$I$ was measured by monitoring the voltage across a load resistor connected in series with the spin-PD. A similar scheme has been used by other groups.[13,14] Finally, we obtain the experimental $F = \Delta I / I_{ph}$.

Two cases were investigated in this study: (1) sidewall illumination and (2) oblique-angle illumination. The plane of incidence is fixed to be parallel to the magnetization vector of the in-plane magnetized Fe electrode. As shown in Fig. 1(b), when $\theta$, which is the angle of a light beam with respect to the axis normal to the sample plane, is set at 90°, the beam is impinged on the cleaved sidewall of the device, that is, sidewall illumination. When $\theta = 60°$, the beam is impinged on the top surface of the device at an oblique angle, that is, oblique-angle illumination. In the case of $\theta = 60°$, the light beam is transmitted through the top metal layers in which it experiences absorption and refraction [Fig. 1(c)]. As a consequence of refraction ($n_{GaAs} = 3.68$),[24] the angle of incidence in the p-GaAs layer is $\theta_{GaAs} = 13.6°$. Additionally, because of the difference between the transmittances for the s- and p- linear polarizations of the metal layers [Fig. 1(c)], the degree of circular polarization of transmitted light, $P = (W_{\sigma+} - W_{\sigma-}) / (W_{\sigma+} + W_{\sigma-})$ is reduced from pure CP ($P_0 = 1$) to elliptic polarization, namely, $P = P_0 \times (T_s/T_p)$. Here, $T_{s(p)}$, which is the transmittance for the s- (p-) linear polarization of the Au/Ti/Fe metal multilayer, varies as function of angle $\theta$, as shown in Fig. 1(c). $W_{\sigma+(-)}$ is the intensity of the right (left) CP light. Secondly, because there is a light component parallel to the magnetization direction of the Fe layer, a magnetic circular dichroism (MCD) contribution will be present in the measurement with oblique-angle illumination. The contribution of elliptic polarization is considered in the modeling, while that of MCD is discussed in Sect. 4. The magnetization characteristics of the top metal layers were also measured by using a SQUID magnetometer. All experiments were carried out at room temperature.

## 3. Model calculation

In order to investigate the relationship among light intensity, photocurrent, and the helicity-dependent component for both sidewall and oblique-angle illuminations, a simulation based on the combination of drift-diffusion and Julliere spin transport models was carried out. The working equations used here are based on the charge and spin transport equations.[25,26] In particular, we have incorporated the CP-dependent photogeneration term in the spin drift-diffusion equation on the basis of the optical selection rules.[5,6] Our point of interest here are the three-dimensional spatial distributions of non-equilibrium electrons and spins in the conduction band of the p-GaAs region, $\Delta n = \Delta n_\uparrow + \Delta n_\downarrow$ and $\Delta s = \Delta n_\uparrow - \Delta n_\downarrow$. Here, $\Delta n_\uparrow$ and $\Delta n_\downarrow$ represent the non-equilibrium densities of spin-up and -down electrons, respectively. Let us state two equations, the dynamics of $\Delta n$ and $\Delta s$, in Eqs. (1) and (2):

$$\frac{\partial \Delta n}{\partial t} = D\nabla^2 \Delta n + \mu E \nabla \Delta n - \frac{\Delta n}{\tau_{rec}} + G, \tag{1}$$

$$\frac{\partial \Delta s(P)}{\partial t} = D\nabla^2 \Delta s(P) + \mu E \nabla \Delta s(P) - \frac{\Delta s(P)}{\tau_{spin}} + G_{spin}(P). \tag{2}$$

Here, $D$ (= 62 cm²/s in p-GaAs) is the electron diffusion coefficient, $\mu$ is the electron mobility [= 2400 cm²/(V·s) ],[27] $E$ is the electric field, $\tau_{rec}$ (= $7.15 \times 10^{-8}$ s)[28,29] is the minority carrier recombination lifetime, $\tau_{spin}$ (= $2.33 \times 10^{-10}$ s)[29] is the spin lifetime, $G$ is the carrier generation rate due to absorbed photons, and $G_{spin}$ is the spin generation rate [$G_{spin} = (0.5 \times G) \cdot P$, $P = \pm 1$].[5,6] $\tau_{rec}$ and $\tau_{spin}$ are estimated on the basis of data cited from Refs. 27-29. In Eqs. (1) and (2), the first term accounts for the diffusion transport, the second term for the drift, the third term for relaxation to the equilibrium, and the last term for photogeneration.

We use the steady-state solution of the form $\Delta n \approx \Delta n(x) \cdot \Delta n(y) \cdot \Delta n(z)$ ( $\Delta s \approx \Delta s(x) \cdot \Delta s(y) \cdot \Delta(z)$ ) in order to simplify Eqs. (1) and (2) into three, one-dimensional (1D) equations. The solution form of the 1D equation [Eq. (3)] is already available in the literature.[30-32]

$$\Delta n(x) = A \exp\left(\frac{+x}{\sqrt{D\tau_{rec}}}\right) + B \exp\left(\frac{-x}{\sqrt{D\tau_{rec}}}\right) + Cg(x) \tag{3}$$

Here, $A$, $B$, and $C$ are constants determined by applying the boundary conditions, and $g(x)$ is a particular solution for a given $G(x)$. The flows $\Delta n$ and $\Delta s$ then pass across the tunnel barrier, through which the photocurrent $I_{ph}$ and the helicity-dependent term $\Delta I$ are defined by Eqs. (4) to (6).

$$I_{ph} = -e \iint \mu E \Delta n(z=0)\, dxdy, \tag{4}$$

$$I_s(P) = -e \iint \mu E \Delta s(z=0,P)\, dxdy, \tag{5}$$

$$\Delta I \approx I_{ph} \frac{\Delta R}{R_{total}} = I_{ph} \left\{ \frac{\frac{1}{2}P_{Fe}[I_s(P=+1)-I_s(P=-1)]/I_{ph}}{1-\frac{1}{2}P_{Fe}[I_s(P=+1)-I_s(P=-1)]/I_{ph}} \right\}. \tag{6}$$

Here, $\Delta R$ denotes a change in the resistance of the spin-PD caused by the switching of helicity, whereas $R_{total} = R_{SPD} + R_{load}$ is the total series resistance of the spin-PD, $R_{SPD}$, and load resistor, $R_{load}$ (described in Sect. 2). Equation (6) has been developed explicitly in the present work on the basis of the Julliere tunneling model, which is presented in Eq. (7). Here, $\Delta R/R$ is the tunnel magnetoresistance ratio (TMR), while $P_1$ and $P_2$ correspond to the spin polarization of the density of states of magnetic metal contacts 1 and 2, respectively. In our case, contact 1 corresponds to the Fe electrode and contact 2 to the conduction band of p-GaAs in which spin-polarized electrons are accommodated. Concretely stated, we plug in the spin polarization of the photogenerated electrons at the γ-AlO$_x$/p-GaAs interface.[25]

$$\frac{\Delta R}{R} = \frac{2P_1P_2}{1-P_1P_2} \tag{7}$$

Note that, in experiments, $R_{load} = R_{SPD}$.[16] Under this condition, the measured helicity-dependent resistance is actually half that of the Julliere TMR, $\Delta R/R_{total} = \frac{1}{2}(\Delta R/R)$. This is reflected in the omission of the factor of 2 in Eq. (6) as compared with Eq. (7). All numerical calculations were carried out by the conventional finite element method and implemented in MATLAB.

A schematic diagram of the simulation geometry is shown in Fig. 2. We start with the case of sidewall illumination [Fig. 2(a)], in which homogeneous illumination within the illuminated area (the *y-z* plane at *x* = 0) is assumed. We ignore the problem associated with illumination boundary. The generation rate $G$ has a spatial profile described by the Beer-Lambert law $G(x) = \alpha(1 - R)I_0 exp(-\alpha x)$. Here, $\alpha$ is the absorption coefficient of GaAs at 785 nm, $R$ the reflectance of the air/p-GaAs interface, and $I_0$ the photon flux of the incident beam in the unit of number of photons per cm$^2$. As such, generation mostly happens near the surface of the sample ($1/\alpha \approx 0.7$ μm). The strength of the electric field in the Schottky depletion region is estimated to be $|E| \approx 2.1 \times 10^5$ V/cm. The surface recombination is taken into account in the boundary condition $J_{surf} = \partial \Delta n/\partial x = S \cdot \Delta n(x = 0)/D$.[30,31] Here, $S$ is the surface recombination velocity, which we will treat as a variable parameter later in Sect. 5.

Shown in Figs. 3(a) and 3(b) are the Δ*n* and Δ*s* profiles in the *x-z* plane, respectively. On the basis of the slopes of the profiles, the direction of electron flows can be inferred. It can be seen that the space contributing to the flow of electrons $J_{e,ph}$ (where $I_{ph} = - e \cdot J_{e,ph}$) is ~30 μm wide (in the *z*-axis) and ~30 μm deep (in the *x*-axis). In contrast, the contributing volume to the flow of spins $J_{e,s}$ (where $I_s = - e \cdot J_{e,s}$) is only ~2 μm wide and ~2 μm deep. This can be understood by considering the difference between the carrier diffusion length ~21 μm and the spin diffusion length ~1.3 μm.[29] Note that the direction of the arrows indicate the direction of the flow of electrons ($J_{e,ph}$) and spins ($J_{e,s}$), which is opposite to that of the actual currents ($I_{ph}$ and $I_s$).

For oblique-angle illumination [Fig. 2(b)], the illuminated area (the *x-y* plane at *z* = 0) is now elliptic, and the generation rate $G$ is expressed in the form $G(z) = \alpha T I_0 exp(-\alpha z/cos\theta_{GaAs})/\cos\theta$, $\theta = 60°$. Here, the transmittance $T$ through the Au/Ti/Fe layers is $2.46 \times 10^{-3}$, and $\theta_{GaAs}$, which is the angle of a light beam inside GaAs with respect to the normal axis, is $\theta_{GaAs} = 13.6°$ [inset of Fig. 1(c)]. Note that the degree of circular polarization is reduced to $P = T_s/T_p = 0.33$. Only the spin axis component parallel to the in-plane magnetization of the Fe layer has to be taken into account. Furthermore, the effect of MCD is not considered in the calculation. The surface recombination due to the $\gamma$-AlO$_x$/p-GaAs interface is assumed to be small.

Shown in Fig. 4 is the $\Delta s$ profile as a function of $z$, while the $\Delta n$ profile is shown in the inset. It can be seen that the peak $\Delta n$ and $\Delta s$ are smaller than those of the sidewall case (Fig. 3), which is mainly due to the small transmittance of the top metal layers. The relatively short diffusion profile of $\Delta s$ (~2 μm) compared with the $\Delta n$ profile (~30 μm) suggests a highly efficient collection of spins, since the excitation is confined near the $z = 0$ plane.

## 4. Results

Shown in Fig. 5(a) is the measured photocurrent plotted as a function of time for sidewall illumination as the helicity of the light is switched. Here, no electrical bias is applied. The photocurrent averaged over observation time is $I_{ph} \approx 10.4$ μA. Clear steplike switching profiles can be seen as the helicity of the light is changed. Moreover, the step profiles become opposite when the magnetization direction is reversed, as expected. That is, the amplitude $\Delta I$ in the helicity-dependent photocurrent is negative (positive) for the $H = +1.35$ kOe ($H = -1.35$ kOe). Here, the average step amplitude appears to be $\Delta I \approx 0.015$ μA.

Shown in Fig. 5(b) is the $F$ - $H$ profile for sidewall illumination together with the magnetization hysteresis curve of the 50-nm-thick Fe layer. It can be seen that, although the Fe layer exhibits remanent magnetization, the $F$ - $H$ profile does not show remanence. Indeed, $F$ becomes measurable only when $|H|$ is greater than 500 Oe; $F = \Delta I/I_{ph} \approx$ 0.14% for $|H| > 500$ Oe.

Similarly, Fig. 5(c) shows the temporal plot of the photocurrent for oblique-angle illumination. No external magnetic field is applied during the measurement. The steplike change due to the helicity-dependent photocurrent term can be seen more clearly, which is inverted when the Fe electrode is magnetized in the opposite direction. In this case, the measured average $I_{ph}$ and $\Delta I$ are ~29.0 and ~0.33 μA, respectively. Shown in Fig. 5(d) is the $F$ - $H$ profile for oblique-angle illumination together with the magnetization hysteresis curve. It can be seen that the $F$ profile closely matches that of the magnetization curve. A measurable $F$ is observed at remanence, $F \approx 1.2$%; this is approximately 10 times higher than that of sidewall illumination and is comparable to that reported for a vertical-type spin-PD.[9,10]

Because light is transmitted through the metal layers for oblique-angle illumination, one may argue that there is some contribution of the MCD component on *F*. If the MCD component is significant in the photocurrent, then *F* should be nearly independent of electrical bias. Shown in Figs. 6(a) and 6(b) are the *F* - *H* profiles for $V = -1$ and $+1$ V, respectively. *F* is nearly unchanged when the reverse bias is applied ($V = -1$ V). We infer that the collection efficiency of spin-polarized photogenerated electrons has already reached saturation at lower biases; therefore, no further increase is observed even though the Schottky depletion region widens and the electric field increases. In contrast, when the forward bias is applied ($V = +1$ V), *F* decreases, reaching $F \approx 0.4\%$. This finding suggests that the Schottky junction is under a nearly flat band condition, as expected in the forward-bias regime. The residual *F* may be attributed to MCD or spin polarized hole transport.[9] Nevertheless, these results further support the claim that the observed high *F* at $V = 0$ V and $-1$ V do indeed come from spin-polarized electron transport.

## 5. Discussion

We try to investigate the role of surface recombination at the cleaved sidewall for sidewall illumination. Shown in Fig. 7 are the *calculated values* of $I_{ph}$, $\Delta I$, and *F* as a function of the surface recombination rate *S* (lower horizontal axis) as well as of the surface recombination time $T_{surf}$ (upper horizontal axis). Here, $T_{surf} = 1/(\alpha \cdot S)$, where $1/\alpha$ is the optical penetration depth in GaAs. $T_{surf}$ can be taken as the average time it takes for excited electrons to travel towards the cleaved edge. It can be seen that, as *S* increases, the photocurrent (a blue solid line) decreases. This is expected since, at high *S* values, excited electrons more likely recombine at the cleaved sidewall rather than accumulate in the depletion region and be transported toward the tunneling barrier. Concretely stated, the onset of an abrupt decrease in $I_{ph}$ occurs at approximately $S \approx 10^4$ cm/s at which $T_{surf} \approx 10^{-8}$ s is comparable to the bulk minority carrier lifetime $\tau_{rec}$. The helicity-dependent photocurrent $\Delta I$ (a red solid line) decreases with increasing *S*, but to a much lesser extent since the reduction starts to take place at approximately $S \approx 10^5$ cm/s when $T_{surf} \approx 10^{-9}$ s becomes comparable to the bulk spin lifetime $\tau_{spin}$. Consequently, an abrupt increase in *F* (dark green line) is primarily associated with the reduced $I_{ph}$. In a nutshell, $\Delta I$ is less affected by surface recombination than the photocurrent $I_{ph}$. For example, for $S = 10^7$ cm/s, over 96% of $I_{ph}$ is lost owing to surface recombination, while only 64% is lost for $I_s$. For *S* values much higher than $10^7$ cm/s, both $I_{ph}$ and $\Delta I$ reach almost constant values, reflecting the fact that, at very high *S* values, the number of carriers that undergo surface recombination is limited by the rate of carrier transport from the bulk towards the edge, the carrier-diffusion-limited regime. Comparing the experimentally measured photocurrent in the side-wall configuration with the calculated $I_{ph}$ curve, a good match can be achieved for $S \approx 10^7$ cm/s, which is comparable to the reported value in the literature.[33] In other words, the charge transport can be explained well by the surface recombination process. On the other hand, the experimental values of $\Delta I$ and $F$ are lower than the calculated values by about 2 orders of magnitude. This suggests that surface recombination is not the predominant cause of the low $F$. Since, spin transport mostly occurs within a 2-3 μm from the cleaved edge, we infer other mechanisms such as magnetic edge curling[34] and microscopic damage to the γ-AlO$_x$ tunnel barrier at the cleaved edge. The observation that $F$ tends to saturate at the external magnetic fields at which magnetization saturate [Fig. 5(b)] suggests that the effect due to the magnetic edge curling is remote. On the other hand, the degradation of the tunnel barrier would give rise to the enhancement of the conduction mismatch[35] and thus reduction in spin transport efficiency through tunneling.

A summary of the model calculation for oblique-angle illumination is shown in Fig. 8 together with experimental data. Here, the data are plotted as a function of the beam position on the sample $x$-$y$ surface. As the beam is moved toward the sample edge (along the $x$-axis), part of the beam impinges on the cleaved sidewall. Two different calculations were carried out: one including the sidewall contribution (dashed lines) and the other excluding the sidewall contribution (solid lines). The former shows nonsymmetric profiles around the sample center, reflecting a relatively large contribution of $\Delta I$ from the sidewall. On the other hand, the latter exhibits symmetric profiles. The experimental data rather follow the latter scenario. The slightly larger experimental data than the calculated $F$ curve (dark green solid line) might be attributed to the contribution from MCD.

## 6. Conclusions

In this work, we have studied the optoelectronic characteristics of a lateral spin photodiode composed of the Fe/γ-AlO$_x$/p-GaAs Schottky junction. Helicity-dependent photocurrent ($\Delta I$) has been measured by illuminating a circularly polarized laser beam with a wavelength $\lambda$ = 785 nm either horizontally on the cleaved sidewall or at an oblique angle on the top metal surface. The plane of incidence is fixed to be parallel to the magnetization vector of the in-plane magnetized Fe electrode. The conversion efficiency $F$, which is defined as the relative value of $\Delta I$ with respect to the total photocurrent $I_{ph}$, has been determined to be $1.0 \times 10^{-3}$ and $1.2 \times 10^{-2}$ for sidewall and oblique-angle illuminations, respectively. Numerical simulation based on the model consisting of drift-diffusion and Julliere spin-tunneling transports has been carried out and simulation results have been compared with the experimental data. Two conclusions have been reached: firstly, the model accounts fairly well for the experimental data without introducing the annihilation of spin-polarized carriers at the γ-AlO$_x$/p-GaAs interface for the oblique-angle illumination; secondly, the model does not fully explain the relatively low $F$ in terms of the surface recombination at the cleaved sidewall in the case of sidewall illumination. We have stated that one of the plausible factors that would be responsible for the degradation near the cleaved sidewall is the microscopic damage to the γ-AlO$_x$ barrier at the cleaved edge. For practical device applications, it would be more advantageous to avoid problems associated with the cleaved edge. This can be circumvented by employing an optical coupling technique to guide a light beam away from the edge, such as by the use of a refracting facet device.[36]

## Acknowledgments

The authors would like to acknowledge the support in part by the Advanced Photon Science Alliance Project from the Ministry of Education, Culture, Sports, Science and Technology (MEXT) and a Grant-in-Aid for Scientific Research (No. 22226002) from the Japan Society for Promotion of Science (JSPS). R.C.R. acknowledges the scholarship from MEXT. N.N. acknowledges the funding from the Ministry of Internal Affairs and Communications (MIC) through the Strategic Information and Communications R&D Promotion Programme (SCOPE) No. 162103004.

**Figure Captions**

**Figure 1** (a) Schematic cross section of the sample structure: from the top, 10-nm-thick Au, 5-nm Ti, 50-nm Fe, 1-nm γ-AlO$_x$, 100-nm Be-doped GaAs epilayer, and a p-GaAs:Zn (001) substrate. (b) Schematic illustration of photocurrent measurement setup. Circular polarization is realized by transmitting a linearly polarized light beam through a linear polarizer (LP) and a quarter-wave plate (QWP). The helicity of the CP light is switched between σ$^+$ and σ$^-$ by manually rotating the QWP. The CP laser beam is focused on the sample surface with a spot size of approximately 400 μm through a lens with the focal length $f$ = 50 cm. (c) Calculated transmittance of the two orthogonal light beam as a function of incidence angle $\theta$. Inset: Schematic of refraction through the top metal layers for oblique-angle illumination with $\theta_{GaAs}$, the angle of a light beam inside the GaAs. (d) Band diagram of the Fe/γ-AlO$_x$/p-GaAs Schottky junction. Upon CP light illumination, spin-polarized electrons (solid grey circles with arrows) and holes (hollow black circles) are generated in the conduction and valence bands, respectively. The built-in electric field in the Schottky depletion region ($|E| \approx 2.1 \times 10^5$ V/cm) separates the carriers and collects the electrons towards the barrier.

**Figure 2** Schematic illustrations for (a) sidewall illumination and (b) oblique-angle illumination, incorporating spatial profiles of photogenerated electrons Δ$n(x)$ and Δ$n(z)$ (black, solid lines), intensity of light that enters the sample (orange, solid lines), and direction of light impinging on the sample (red, waves). The origin of the rectangular coordinate system is set at the γ-AlO$_x$ (white sheets)/p-GaAs (gray blocks) interface. The parameters $\alpha$ and $\theta$ are the absorption coefficient and angle of incidence with respect to the surface normal, respectively. Reddish zone shows the main light absorption region characterized by the 1/e decay of light intensity, with e being the natural logarithm.

**Figure 3** (a) Two-dimensional profile of Δ$n$ in the $x$-$z$ plane. $J_{e,diff}$ is the flow of electrons due to diffusion, $J_{e,surf}$ is the flow of electrons due to surface recombination at the $x$ = 0 plane, and $J_{e,ph}$ is the flow of electrons that are collected at the $z$ = 0 plane. (b) Two-dimensional profile of Δ$s$ in the $x$-$z$ plane. $J_{e,s}$ is the flow of spins collected at the $z$ = 0 plane.

**Figure 4** Profile of $\Delta s$ as a function of $z$. $J_{e,s}$ is the flow of spins collected at the $z = 0$. Inset: Profile of $\Delta n$ as a function of $z$. $J_{e,ph}$ is the flow of electrons collected at the $z = 0$.

**Figure 5** (a) Temporal profiles of photocurrent obtained in the sidewall illumination setup for right ($\sigma^+$) and left ($\sigma^-$) CP laser beams under two opposite external magnetic fields $H = \pm 1.35$ kOe. The applied field is assumed to saturate the in-plane magnetization of the Fe layer. No electrical bias is applied to the sample. (b) Helicity-dependent photocurrent (in the form of $F$) as a function of applied magnetic field obtained in sidewall illumination setup. The magnetization hysteresis loop is also shown. No electrical bias is applied. (c) Temporal profiles of photocurrent obtained in the oblique-angle illumination setup for right ($\sigma^+$) and left ($\sigma^-$) CP laser beams. Fe layers are magnetized parallel (+Rem) or antiparallel (−Rem) along the incident plane ($x$-axis in Fig. 2) prior to the experiments. No electrical bias is applied to the sample. (d) Hysteresis curve of helicity-dependent photocurrent (in the form of $F$) as a function of applied magnetic field obtained in oblique-angle illumination setup. The magnetization hysteresis loop is measured under oblique-angle magnetic fields (60 ° with respect to the surface normal) is also shown. No electrical bias is applied. Note that the axis scales are different for the plots, especially between (a) and (c).

**Figure 6** Hysteresis curves of spin conversion efficiency obtained in the oblique-angle illumination setup for two different electric bias (a) $V = -1$ V and (b) $V = +1$ V. Magnetization hysteresis curves are also shown for comparison.

**Figure 7** Calculated photocurrent $I_{ph}$ (total photocurrent), helicity-dependent component $\Delta I$ (the difference between two opposite spin components), and conversion efficiency $F$ ($\equiv \Delta I/I_{ph}$) as a function of the surface recombination velocity $S$ in the case of sidewall illumination. Surface recombination is assumed to take place at the cleaved ($x = 0$) surface.

**Figure 8.** (a) Experimental photocurrent $I_{ph}$ (blue diamonds), helicity-dependent component $\Delta I$ (red squares), and conversion efficiency $F$ (dark green triangles) as a function of beam position obtained from the oblique-angle illumination setup. Calculated

data are also shown by solid and dashed lines, which are calculated assuming no contribution of the sidewall and assuming contribution of the sidewall under light illumination, respectively. No electric bias is applied to the sample. (b) The sample width and the diameter of the laser beam spot are 900 and 400 μm, respectively.

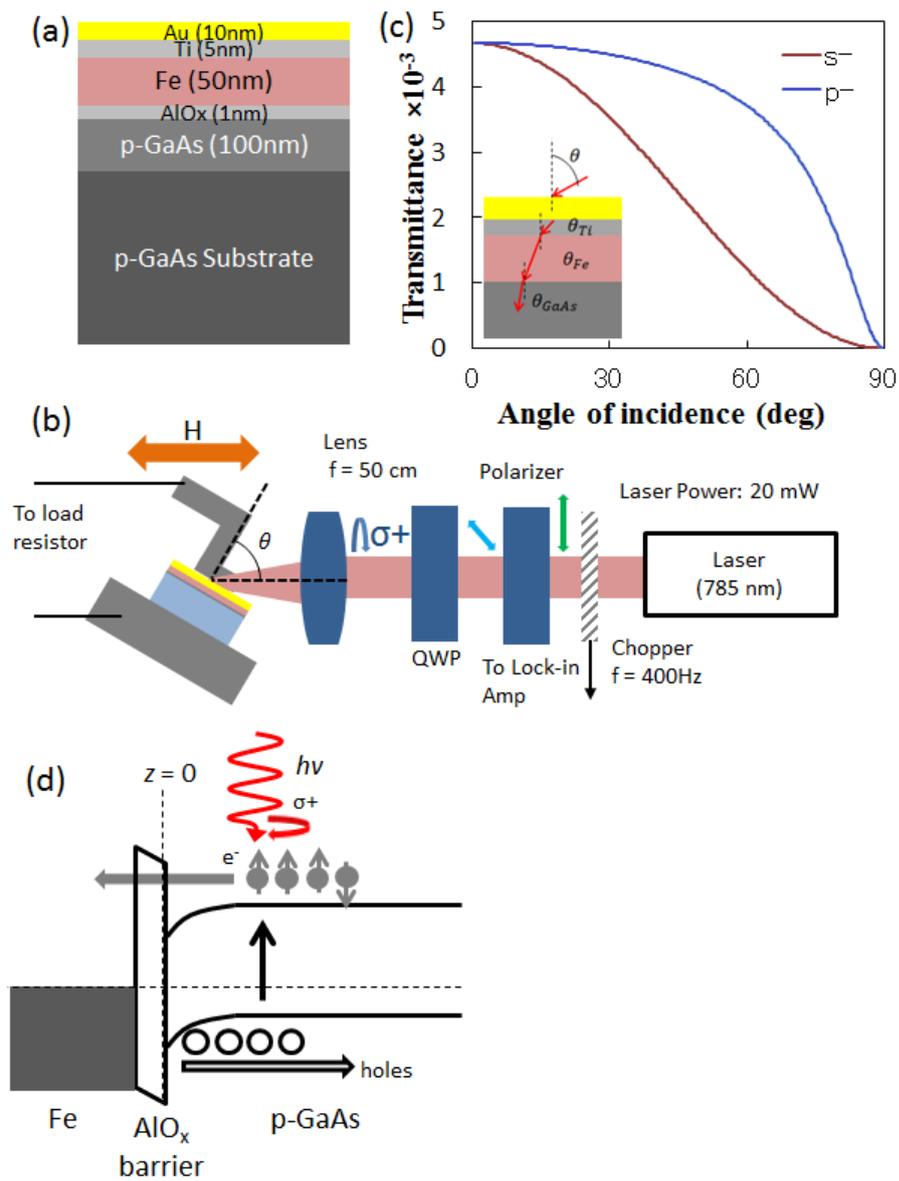

Fig. 1. (Color online)

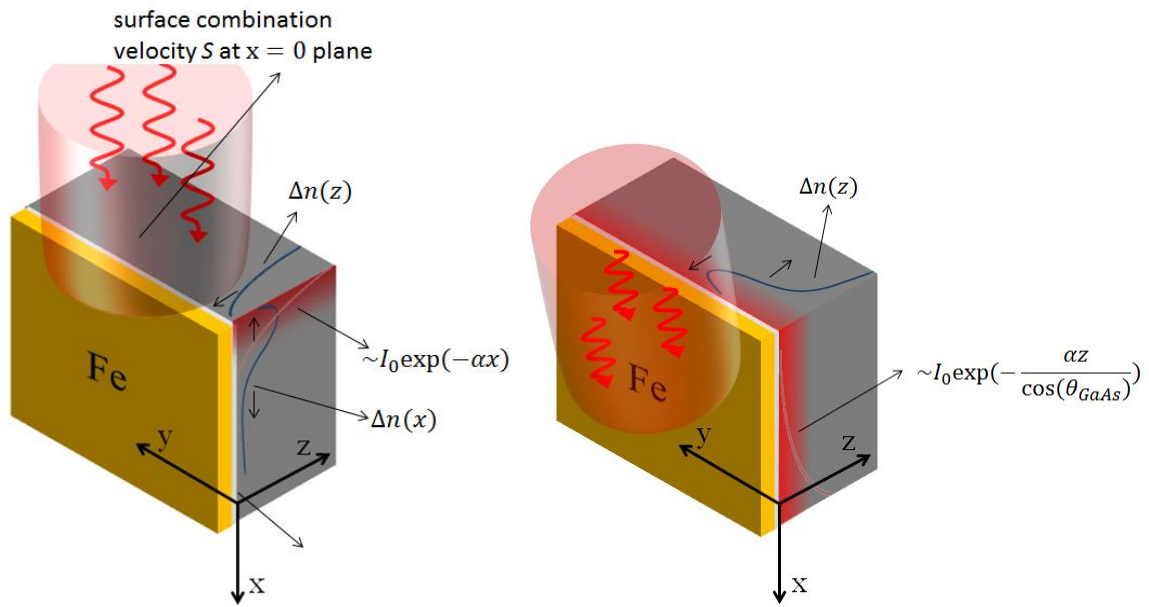

Fig. 2. (Color online)

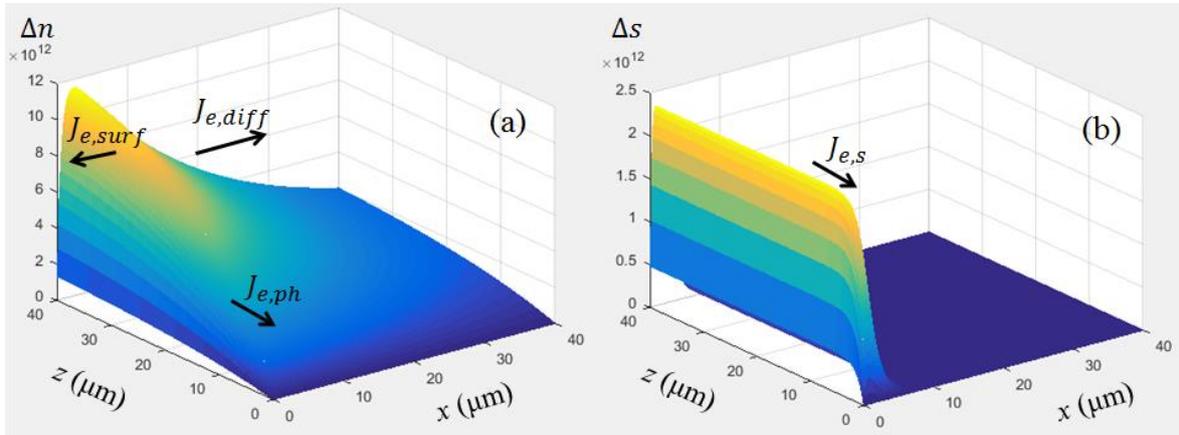

Fig. 3. (Color online)

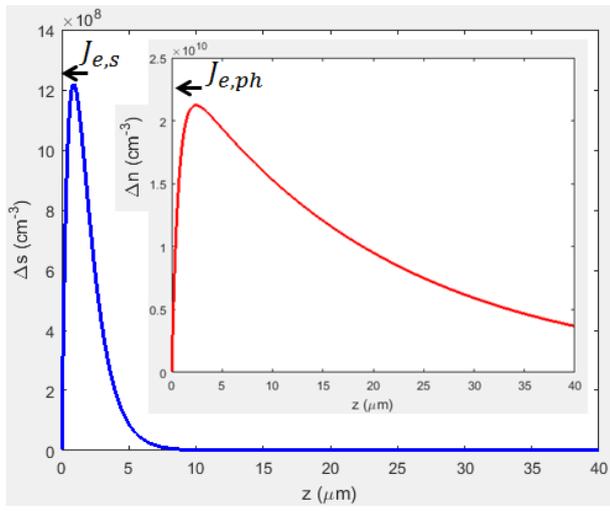

Fig. 4. (Color online)

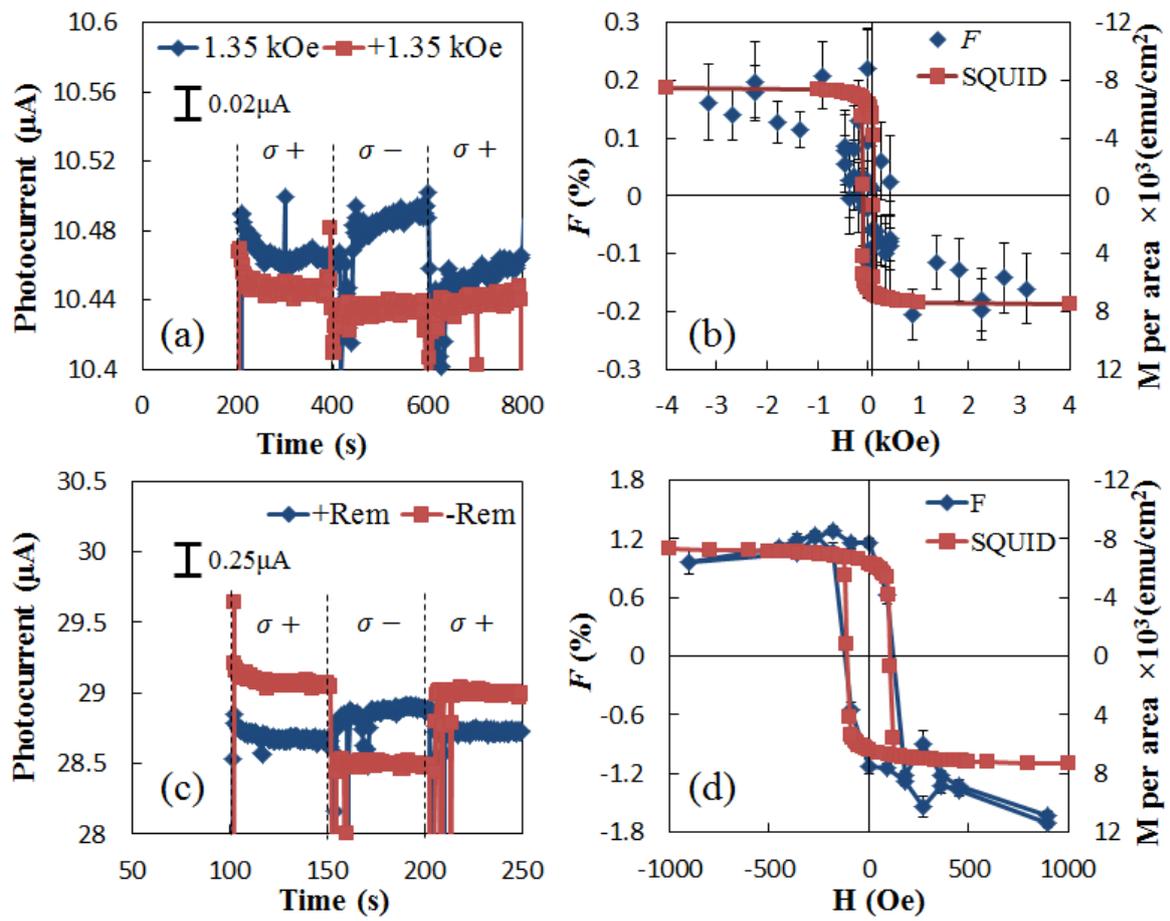

Fig. 5. (Color online)

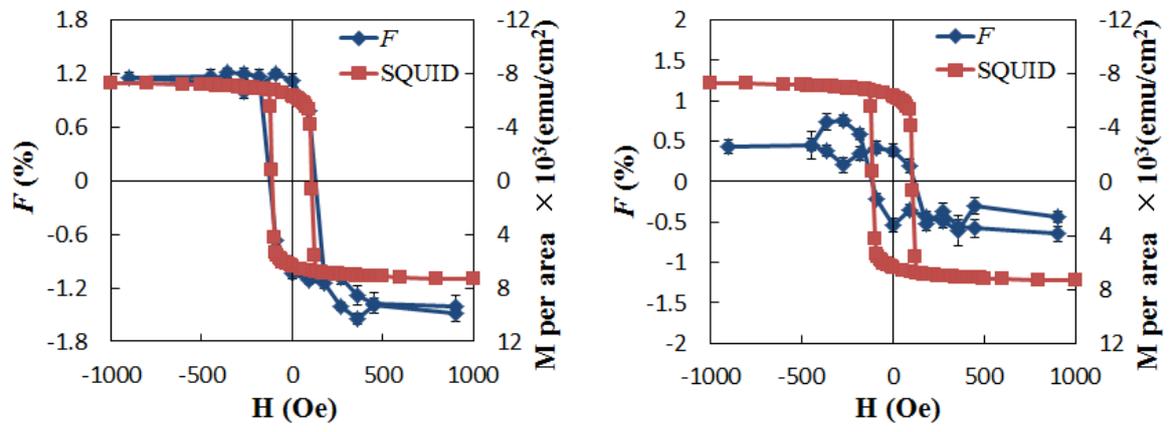

Fig. 6. (Color online)

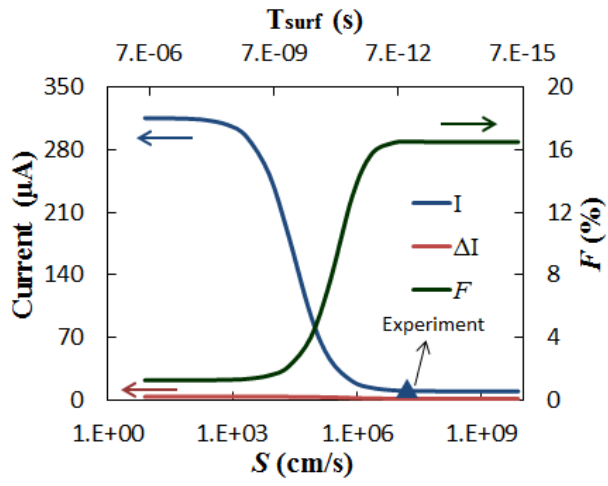

Fig. 7. (Color online)

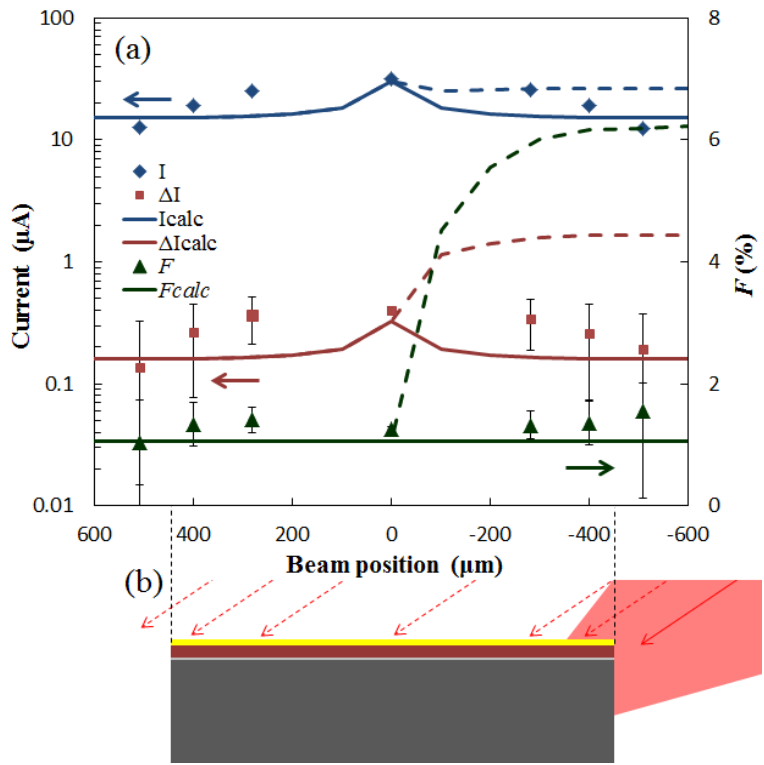

Fig. 8. (Color online)